\documentclass[superscriptaddress,nobibnotes,amsmath,amssymb,notitlepage,twocolumn,prl]{revtex4-2}
\usepackage{mathptmx,bm}
\usepackage{graphicx}
\usepackage[normalem]{ulem}
\usepackage[caption=false]{subfig}
\usepackage{hyperref}
\hypersetup{colorlinks=true, linkcolor=blue, citecolor=blue, urlcolor=blue}
\usepackage{booktabs,multirow}
\graphicspath{{./Fig/}}

\newcommand{\beq}[1]{\begin{equation}\label{#1}}
\newcommand{\eep}{\;.\end{equation}} 
\newcommand{\eec}{\;,\end{equation}} 
\newcommand{\eeq}{\end{equation}}    

\DeclareMathAlphabet{\mathcal}{OMS}{cmsy}{m}{n} 
\renewcommand{\H}{\mathcal{H}}   

\renewcommand{\vec}[1]{{\bf #1}}

\newcommand{\rv}{\vec{r}} 

\let\originalsection\section
\newcommand*{\headingSectionPRL}[1]{\belowpdfbookmark{#1}{#1}{\underline{\textit{#1}}} ---}
\let\section\headingSectionPRL

\newcommand{\HarvardPhysics}{Department of Physics, Harvard University, Cambridge, Massachusetts 02138, USA}
\newcommand{\HarvardSeas}{John A.~Paulson School of Engineering and Applied Sciences, Harvard University, Cambridge, Massachusetts 02138, USA}
\newcommand{\HarvardQSE}{Quantum Science and Engineering, Harvard University, Cambridge, Massachusetts 02138, USA}

\begin{document}

\title{Universal Symmetries in Twisted Moiré Materials}

\author{Mohammed M.~Al Ezzi}
\email{alezzi@seas.harvard.edu}
\affiliation{\HarvardSeas}

\author{Albert Zhu}
\affiliation{\HarvardQSE}

\author{Daniel Bennett}
\affiliation{\HarvardSeas}

\author{Daniel T.~Larson}
\affiliation{\HarvardPhysics}

\author{Efthimios Kaxiras}
\affiliation{\HarvardSeas}
\affiliation{\HarvardPhysics}


\begin{abstract}
\normalsize
Two-dimensional multi-layer materials with an induced 
moiré pattern, either due to strain or relative twist between layers, provide a versatile platform for exploring strongly correlated and topological electronic phenomena. While these systems offer unprecedented tunability, their theoretical description remains challenging due to their complex atomic structures and large unit cells. A 
notable example is twisted bilayer graphene, where even the relevant symmetry group remains unsettled despite its critical role in constructing effective theories. Here, we focus on twisted bilayer graphene and use a combination of analytical methods, molecular dynamics simulations, and first-principles calculations to show that 
twisted atomic configurations with distinct microscopic symmetries converge to a universal interlayer structure that governs the low-energy physics. This emergent universality provides a robust foundation for symmetry-respecting models and offers insight into the role of commensurability in real twisted moiré systems.
\end{abstract}

\maketitle 
\section{Introduction}
Multi-layered two-dimensional materials are revolutionizing condensed matter physics by providing a platform of extraordinary tunability to explore a wide variety of quantum phases \cite{andrei2021marvels}. 
In these systems, 
moir\'e patterns with a tunable length scale much greater than the interatomic distance are produced by adjusting either the relative strain or twist angle between layers (referred to as ``twistronics"~\cite{carr2017twistronics}).
The interplay between the resulting moiré-scale potential and that of the atomic-scale crystal structure of each individual layer can lead to  
correlated electronic behavior, including superconductivity~\cite{cao2018unconventional,yankowitz2019tuning,xia2025superconductivity}, ferromagnetism~\cite{sharpe2019emergent}, correlated insulators \cite{cao2018correlated}, nematicity~\cite{cao2021nematicity}, the quantum anomalous Hall effect~\cite{serlin2020intrinsic}, and fractional Chern insulators~\cite{cai2023signatures,lu2024fractional}, first realized in the context of moiré materials.
Other factors that affect the properties of moiré materials include layer number, temperature, carrier density, and external probes such as electric and magnetic fields. 
From a modeling perspective, the presence of two or more competing potentials with different length and energy scales necessitates a large supercell to capture the relevant features, which can involve tens of thousands of atoms for systems of interest (small strain or small twist angle) \cite{carr2020electronic}.
This makes it extremely challenging to provide an accurate description of such systems based on their atomic structure.  
The atomic-scale structure, however, is the critical element that determines the symmetries of effective theories which try to capture the physics at a more abstract level. 

In the canonical twisted moiré system,  
magic-angle twisted bilayer graphene (tBLG), 
the point group symmetry 
remains an open and critical question. 
Geometric considerations suggest that the system possesses $D_3$ symmetry~\cite{yuan2018model,kang2018symmetry}, while experimental observations, such as the absence of a gap at the charge neutrality point (CNP), suggest $D_6$ symmetry~\cite{po2018origin,zou2018band}. 
The absence of the gap alone is {\it not} sufficient to determine the symmetry because the band crossings at the CNP might occur at different points of the moiré Brillouin zone (mBZ). Recent measurements using quantum twisting microscopy \cite{ilani2025interacting} have revealed a band crossing at the $\Gamma$ point. This is consistent with interacting electron calculations \cite{datta2023heavy}, in sharp contrast with the non-interacting electron picture which yields band crossing at the $K$ point (corner of the hexagonal mBZ). 
The point group symmetry determines the nature of effective lattice models:
$D_3$ symmetry permits a minimal two-band model per valley without topological obstruction~\cite{koshino2018maximally,kang2018symmetry}, while $D_6$ symmetry introduces a fragile topological obstruction that requires larger Hilbert spaces~\cite{po2018fragile,po2019faithful}.
Clarifying the relevant underlying symmetries of moiré materials is therefore essential for modeling and understanding correlated phases.

In this Letter we show that the effective symmetry of twisted bilayers is not determined by the exact geometry of the full, commensurate supercell, but rather by the symmetry that emerges due to uniform sampling of the possible local stacking environments that govern interlayer interactions and ultimately dictate the atomic relaxation and electronic structure.
We systematically construct twisted samples with different point group symmetries and examine how their features evolve as a function of twist angle, using the configuration space formalism \cite{Cazeaux2017,carr2018relaxation}.
Taking tBLG as an example, our approach reveals that, for a twist angle smaller than $10^{\circ}$, the atomic and electronic degrees of freedom exhibit universal behavior governed by the same symmetry group, regardless of geometric differences between commensurate structures.
This is demonstrated through large-scale first-principles density functional theory (DFT) calculations and molecular dynamics (MD) simulations. 
Furthermore, we argue that incommensurate structures, 
which can be considered as compositions of many slightly different
commensurate structures, should display the same effective symmetry. 
This perspective provides a unified framework for understanding symmetry in tBLG and offers a pathway towards the design of more accurate and physically motivated lattice models.

\begin{figure}[t!]  
\centering
\includegraphics[width=1.0\columnwidth]{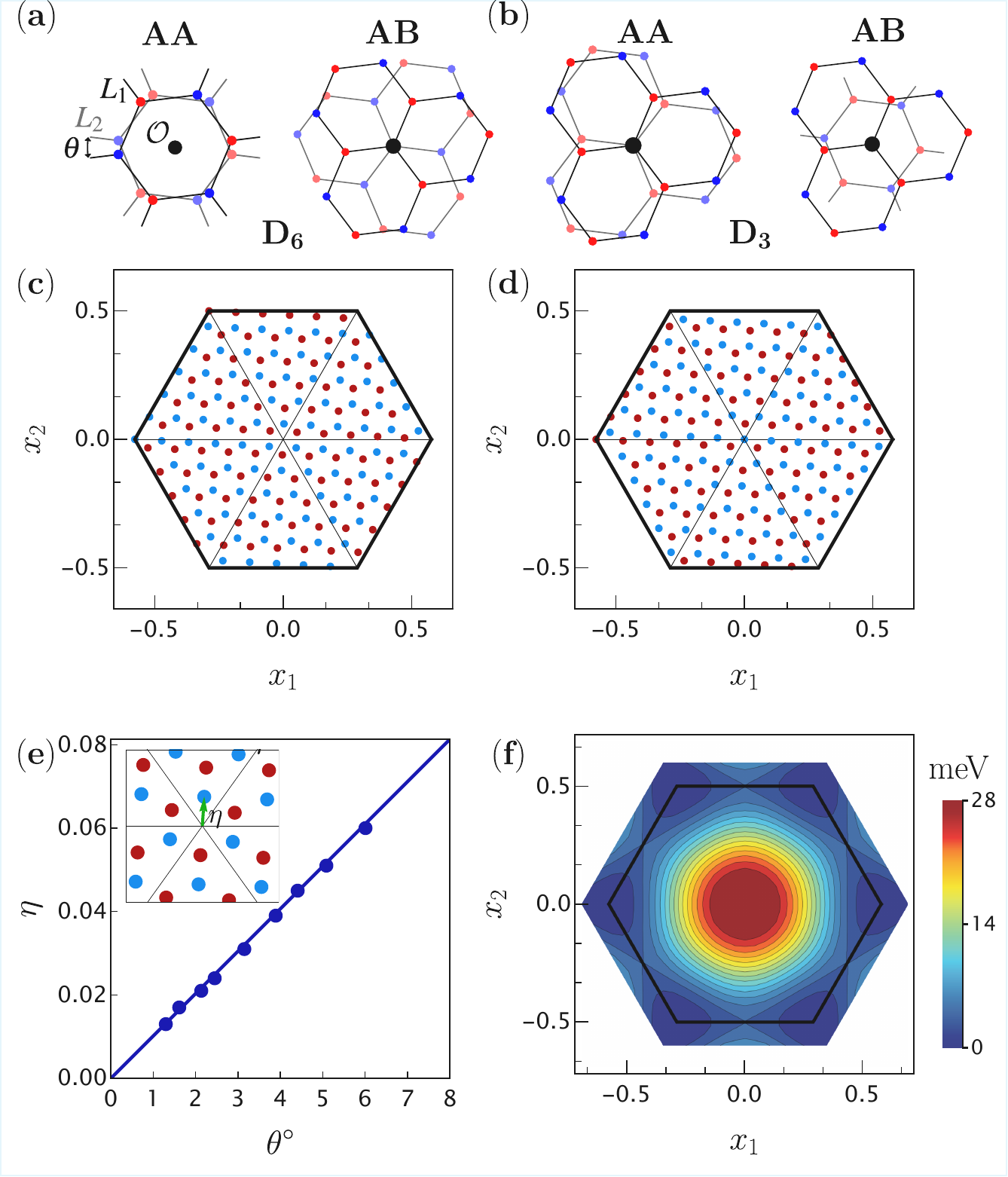}
\caption{{\bf(a-b)} Representative configurations for different choices of stacking (AA or AB) and rotation centers (black dots), resulting in supercells with $D_6$ symmetry in (a) and $D_3$ symmetry in (b). Note that in (a) the $D_6$ symmetry axis for AB stacking arises at a different location within the commensurate cell. 
{\bf(c-d)} CS grids with $D_6$ and $D_3$ symmetry, respectively, for unrelaxed tBLG structures with $\theta = 6.0^{\circ}$. 
The red and blue color indicates the CS grid for the two sublattices.
{\bf (e)} Magnitude of the rigid shift $\eta$ between $D_6$ and $D_3$ CS grids as a function of twist angle $\theta$, showing the linear behavior of Eq.~(\ref{equ:eta}).
The points are extracted from unrelaxed structures.
The inset shows a close-up of the origin in (c), with $\eta$ indicated by the green arrow.
{\bf (f)} GSFE for tBLG, obtained from MD simulations. 
The energy landscape reflects stacking-dependent interactions, with a peak at the center corresponding to AA stacking. All lengths are given in units of the primitive lattice constant of the monolayer, making $\eta$ a dimensionless quantity.
}
\label{fig1}
\end{figure}

\section{Geometric features}
The physical properties of real samples are dictated by the geometries realized experimentally. 
Since the fabrication process of moiré  materials cannot be perfectly controlled, randomness in the initial stacking and rotation centers is introduced~\cite{lau2022reproducibility}. 
Alternatively, computational atomistic models can provide insight into commensurate structures. 
In untwisted bilayer graphene there are two high-symmetry stacking arrangements: AA, where the layers are vertically aligned, and AB, where only one sublattice of carbon atoms is vertically aligned with the opposite sublattice in the other layer
with each stacking possessing several high-symmetry rotation centers. 
Combinations of these stackings and rotation centers can be used to generate high-symmetry twisted commensurate structures.
Selecting as rotation centers either a point corresponding to vertically aligned atoms or a point corresponding to the center of aligned hexagons, or a combination of the two, yields four representative configurations.
The symmetry groups of the resulting twisted configurations, $D_6$ or $D_3$, are summarized in Table~\ref{tabS1} of the Supplemental Material~\cite{SM}, while the atomic configurations are illustrated in Fig. \ref{fig1} (a) and (b).

The two moiré superlattices with $D_3$ symmetry, although generated from untwisted bilayers with AA and AB stacking, differ only in the origin of the moir\'{e} cell.
The same is true for the two configurations with $D_6$ symmetry.
Thus, for a given commensurate twist angle, there are two distinct high-symmetry twisted bilayer configurations with one moiré pattern per unit cell: one with $D_3$ symmetry and one with $D_6$ symmetry. 
Previously, these configurations were identified as distinct structures arising from different but complementary twist angles~\cite{mele2010commensuration,lopes2012continuum}. 
Viewing them as arising from different stacking configurations and rotation centers with a common twist angle offers a clear and
consistent basis for comparing the atomic structures with $D_3$ and $D_6$ symmetries.
There are also commensurate structures of lower symmetry which we discuss later.  

Our MD calculations, discussed below, show that the total energies of the two high-symmetry configurations with the same twist angle and 
moiré-scale periodicity
are identical within the numerical precision of the method, implying equal likelihood of experimental realization. 
Consequently, when building lattice models both geometries must be considered on equal footing, making it essential to determine the correct electronic symmetry for accurately describing correlated phases in tBLG and other moiré materials.

\section{Symmetry of the interlayer hamiltonian}
The hamiltonian of tBLG can be written as 
\beq{eq:H}
\H = \H_1 + \H_2 + \H_{\perp}, 
\eeq
with $\H_1$, $\H_2$ being the {\em intralayer} hamiltonians for layers 1 and 2, respectively, and $\H_{\perp}$ representing the {\em interlayer} coupling. 
The physics of tBLG is 
governed by the moiré-scale $\H_{\perp}$, 
which
provides the proper
framework for understanding the relationship between twisted configurations of different symmetry. 
For non-interacting layers
($\H_{\perp}=0$), 
the spectrum would be that of two copies of monolayer graphene.
For interacting, twisted layers, the intralayer physics of monolayer graphene is modulated by
$\H_{\perp}$ 
which is, at least approximately periodic on the moir\'{e} scale.
Here we numerically evaluate how well the approximate symmetries of
$\H_{\perp}$ 
manifest in the properties of a twisted structure.

A useful framework for this investigation is the ``configuration space'' (CS)
\cite{Cazeaux2017,carr2018relaxation}, which employs the concept of local atomic environments. 
Setting aside temporarily the effect of atomic relaxation, 
the possible local environments of a given carbon atom are determined by the location of the nearest corresponding ({\em i.e.} same sublattice) atom in the opposite layer, 
with all in-plane vectors that connects such pairs of atoms constituting the CS,
a periodic vector space with lattice constant $a$, which is isomorphic to the primitive untwisted unit cell.
For a commensurate supercell the CS grid contains one point for each atom in a single layer, and together the points form a honeycomb lattice with lattice constant $a/\lambda$, where $\lambda=a/2\sin(\theta/2)$ is the moir\'{e} period. 
Examples of the CS grid for structures with $D_6$ and $D_3$ symmetry for twist angle 6.0$^\circ$ are shown in Figs.~\ref{fig1}(c) and (d), respectively. Both grids clearly show the underlying $D_6$ and $D_3$ symmetry of the corresponding tBLG supercells. 

The $D_6$ and $D_3$ CS grids only differ by a rigid shift which we define as $\eta$, highlighted in the inset of Fig \ref{fig1}(e); in this figure, we show 
$ \eta$ as a function of twist angle, which decreases linearly and approaches zero at small angles. 
As the twist angle decreases, the number of atoms per moiré supercell increases, while the area of CS remains fixed. 
This leads to a higher density of points in CS, hence a smaller shift $\eta$ between the $D_3$ and $D_6$ CS grids. The linear behavior can be analytically understood, by noting that the maximum value for the shift 
 is the distance in the CS grid between a grid point and the hexagon center:
\begin{equation}
\eta_{\rm max} = \frac{1}{\sqrt{3}}\frac{a}{\lambda} = \frac{2}{\sqrt{3}}\sin\left(\frac{\theta}{2}\right)
\approx \frac{\theta}{\sqrt{3}}.
\label{equ:eta}
\end{equation}
To relate this geometric difference in CS to energy, we compute the generalized stacking fault energy (GSFE)~\cite{kaxiras1993free}, shown in Fig.~\ref{fig1}(f), which is used to calculate global properties of the twisted bilayers. 



\begin{figure}[!t]
\includegraphics[width=1.0\linewidth]{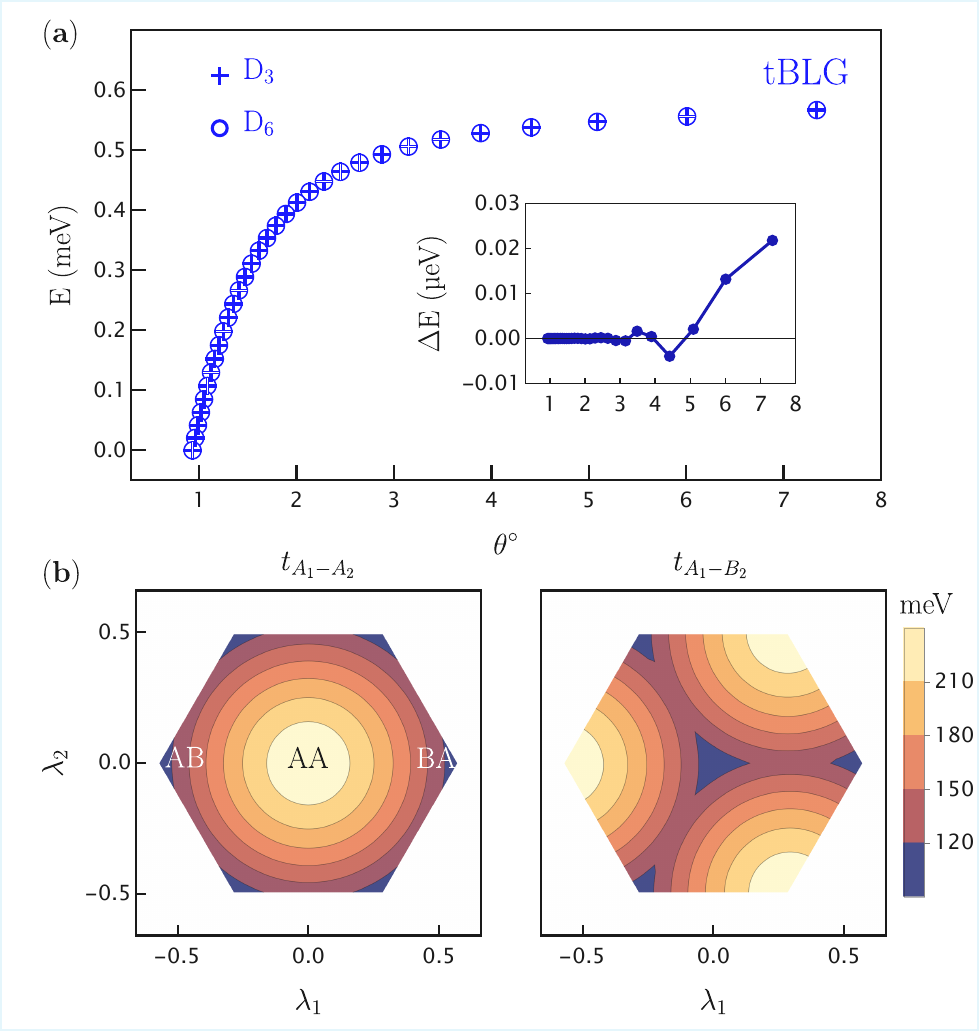}
\caption{Cohesive energy and moiré potential.
{\bf (a)} Energy per atom for tBLG as a function of twist angle for both $D_3$ and $D_6$ configurations. 
Energies are measured relative to the energy of the configuration with the smallest twist angle. 
The inset shows the difference in energy per atom for the $D_3$ and $D_6$ configurations of tBLG as a function of twist angle. 
{\bf (b)} Moiré potential for tBLG for $\theta=1.05^\circ$ with $D_3$ symmetry, showing intra-sublattice coupling $t_{A_1-A_2}$ (left) and inter-sublattice coupling $t_{A_1-B_2}$ (right). 
The profile of the moiré interlayer potential for $D_3$ is identical to that for $D_6$ configurations (not shown),  
in terms of both symmetry and magnitude.
}
\label{fig2}
\end{figure}

\begin{figure}[!t]
\centering
\includegraphics[width=1.0\linewidth]{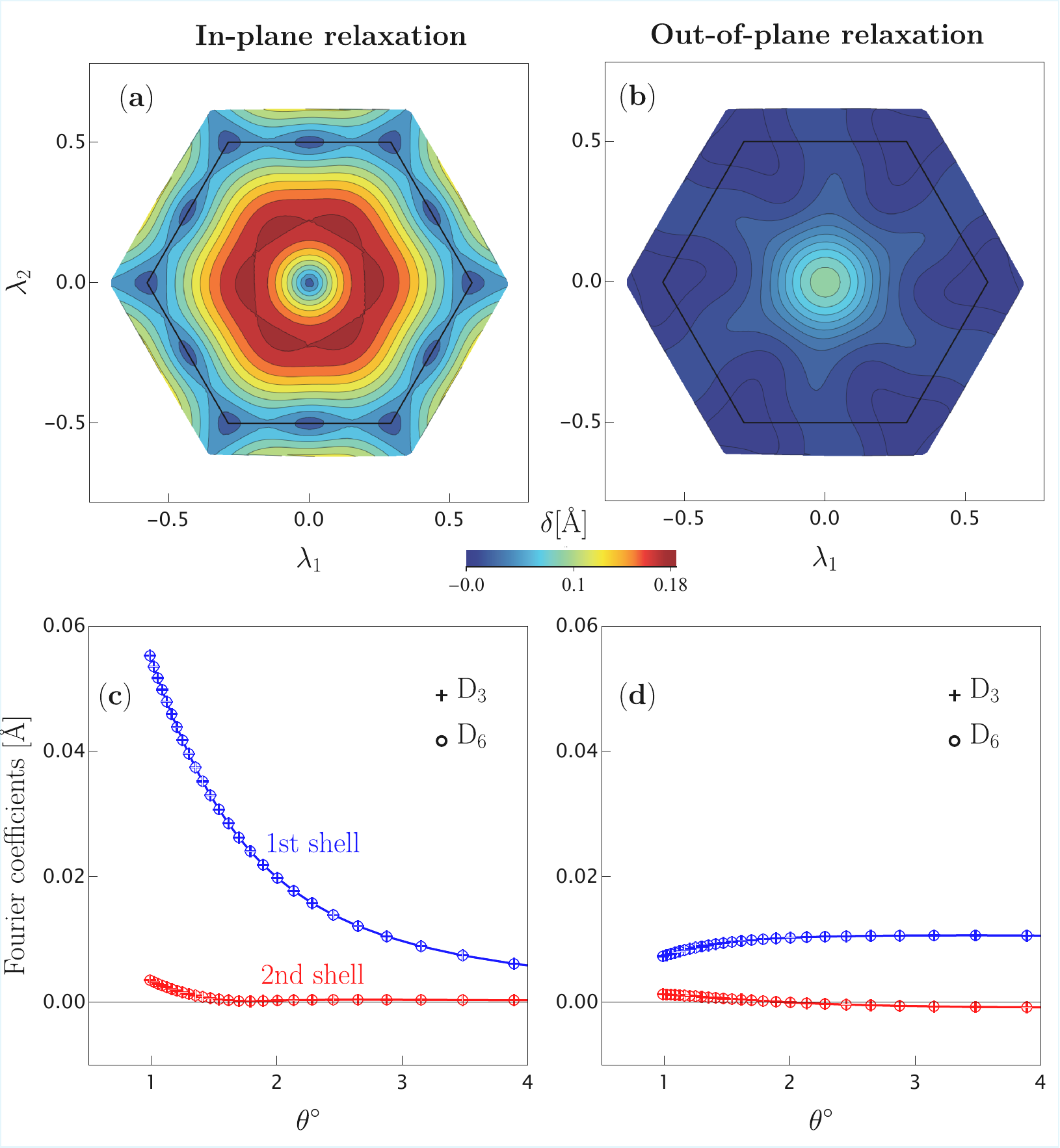}
\caption{
Atomic displacement fields in tBLG.
The displacement fields for $D_3$ and $D_6$ are identical; accordingly, we show the field for only one configuration, and the first two Fourier components for the two configurations.
Magnitude of the displacement field at a twist angle $\theta = 1.05^{\circ}$, are shown for 
{\bf (a)} in-plane, and 
{\bf (b)} out-of-plane displacements.
{\bf (c–d)} Magnitude of the Fourier components $|\mathbf{f}_n|$ for the first two shells (defined by vectors $|\mathbf{g}_n|$ of smallest magnitude in Eq.~\eqref{eq:displacement}) for the $D_3$ and $D_6$ displacement fields, corresponding to {\bf (c)} in-plane and {\bf (d)} out-of-plane relaxations. 
}
\label{fig3}
\end{figure}


\section{Cohesive energy and moiré potential}
While CS offers a powerful way to analyze the local structure of the interlayer hamiltonian, it does not capture global atomic and electronic effects.
To obtain full-system insight, we start by performing molecular dynamics simulations of twisted bilayer samples. 
For the physical picture developed above to hold, the total energies of the $D_3$ and $D_6$ configurations should be nearly identical.
Fig.~\ref{fig2}(a) shows the cohesive energy per atom for the $D_3$ and $D_6$ bilayer graphene structures,
which are indistinguishable at the meV scale 
(the energy difference between the two configurations is numerically zero for $\theta < 4^{\circ}$ and remains below $0.03$ $\mu$eV for $\theta < 10^{\circ}$, see inset).
For other twisted bilayer materials, like WSe$_2$, we find similar results: configurations arising from different twist centers yield identical cohesive energy. This suggests that experimentally realizable structures will exhibit the same behavior regardless of the microscopic twist center.

To further validate the equivalence of $\H_{\perp}$ in the $D_3$ and $D_6$ configurations, we compare their real-space moiré potentials \cite{lopes2007graphene}. 
Fig.~\ref{fig2}(b) shows the strength of the interlayer intra-sublattice interaction  ($t_{A_1-A_2}$, left panel) and interlayer, inter-sublattice interaction ($t_{A_1-B_2}$, right panel) for the $D_3$ configuration. 
The spatial profiles exhibit the same symmetry and magnitude for both $D_3$ and $D_6$, reinforcing the conclusion that the interlayer hamiltonian is effectively identical for $D_3$ and $D_6$ at small twist angles. 
Identical interlayer interactions should manifest in identical physical observables, as we discuss next.

\section{Atomic displacement fields}
In layered two-dimensional materials, van der Waals interactions favor low-energy local stacking configurations, causing the atoms in tBLG to relax into a minimum energy geometric pattern consisting of large AB/BA domains separated by domain walls which intersect at AA domains, that in turn form a triangular lattice \cite{nam2017lattice,carr2018relaxation, zhang2018structural, ezzi2024analytical}.
To examine the effect of symmetry on atomic relaxation, we performed MD simulations using {\sc lammps} \cite{THOMPSON2022108171,wen2018dihedral,naik2019kolmogorov,SM}. 
Figs.~\ref{fig3}(a) and (b) show the magnitude of the in-plane and out-of-plane displacement fields for one layer in the $D_3$ configuration with a twist angle of 1.05$^\circ$. (Due to the symmetry of the layers, the other layer exhibits the same pattern with the opposite orientation.) 
We find that the relaxation fields for the $D_6$ configuration are identical within numerical precision to those of $D_3$, confirming that the interlayer hamiltonian $\H_{\perp}$ is effectively the same for small twist angles. 
To quantify this, we specify the relaxed atomic positions by: 
\begin{equation}
\mathbf{r} = \mathbf{r}_0 + \delta \mathbf{f}(\mathbf{r}_0), \quad
\delta \mathbf{f}(\mathbf{r}_0) = \sum_n \mathbf{f}_n e^{i \vec{g}_n \cdot \mathbf{r}_0}
\label{eq:displacement}
\end{equation}

\noindent where $\mathbf{r_0}$ is the unrelaxed (rigid) atomic position, $\delta \mathbf{f}(\mathbf{r_0})$ is the displacement field, $\mathbf{f_n}$ are Fourier coefficients and $\mathbf{g_n}$ are the moiré reciprocal lattice vectors. 
In Figs~\ref{fig3}(c) and (d) we show the magnitude of Fourier components $|\mathbf{f}_n|$ of the $D_3$ and $D_6$ displacement fields, for the first two shells, defined by vectors $\mathbf{g}_n$ with the smallest non-zero magnitude in Eq.~\eqref{eq:displacement}. 
Not only is the difference in the magnitudes negligible between the $D_3$ and $D_6$ cases, but their symmetry is consistent with that of the $D_6$ point group
(as proven by the actual values of the vectors $|\mathbf{f}_n|$), 
in which the in-plane displacement fields are purely rotational~\cite{ezzi2024analytical}.
It is worth emphasizing that both configurations retain their original total symmetry after relaxation, contrary to previous reports~\cite{angeli2018emergent}. Even though the total atomic field $\mathbf{r}$ inherits the symmetry of the initial stacking 
($D_3$ or $D_6$), $\delta \mathbf{f}(\mathbf{r_0})$ is identical for both $D_3$ and $D_6$ configurations. 
This highlights the effectiveness of symmetry-based analytical models in describing complex atomic relaxation in moiré systems.

\begin{figure*}[t]
\centering
\includegraphics[width=1.0\textwidth]{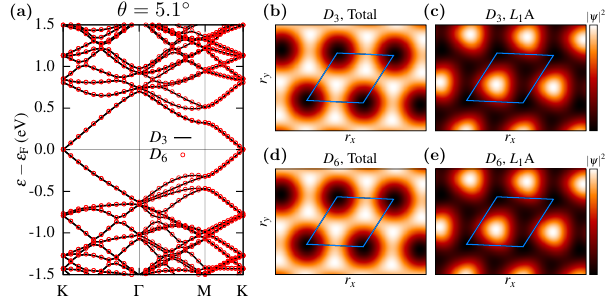}
\caption{
Symmetry of electronic properties in tBLG.  
{\bf (a)} Band structure at a twist angle $\theta = 5.1^{\circ}$ for the $D_3$ (black lines) and $D_6$ (red dots) configurations. 
Representative wave function of the first conduction band at the $\Gamma$ point for the $D_3$ configuration: 
{\bf (b)} magnitude of the total wave function; 
{\bf (c)} projection onto the A sublattice of the first layer ($L_1$). 
{\bf (d–e)} Same as in {\bf (b–c)} for the $D_6$ configuration. 
}
\label{fig4}
\end{figure*}

\section{Symmetry of electronic features}
To study the effect of $\H_{\perp}$ on the electronic properties, a full quantum mechanical treatment is required. 
We compute the band structures and representative wavefunctions for both $D_3$ and $D_6$ configurations for tBLG
with density functional theory (DFT) using the \textsc{siesta} code \cite{soler2002siesta,SM}, with a localized numerical atomic orbital basis set optimized for graphene \cite{bennett2025accurate}. 
Fig.~\ref{fig4}(a) shows the resulting band structures for $D_6$ and $D_3$ configurations 
at $\theta = 5.1^\circ$. 
The two spectra are indistinguishable, providing further evidence that $\H_{\perp}$ is essentially the same in both configurations.
In Figs.~\ref{fig4}(b) and (c)
we show representative wavefunctions at the $\Gamma$ point for both systems:
the total wavefunction magnitude and its projection on one sublattice of one layer  
are indistinguishable between the two configurations. Furthermore, analyzing their symmetry character (Table \ref{tabS2}, in the Supplemental Material \cite{SM}) confirms that the wavefunctions of both configurations exhibit $D_6$ symmetry \cite{angeli2018emergent}, 
thus demonstrating that the universal $D_6$ symmetry of the atomic relaxation  
extends to the electronic degrees of freedom. 

\section{Lower symmetry structures}
Thus far we have focused on high-symmetry commensurate structures. However, commensurate structures with reduced symmetry also exist. These can be generated by starting from one of the high-symmetry stackings and applying a rigid translation to either one or both layers before or after twisting. 
These lower-symmetry structures are each characterized by a CS grid that differs from the $D_3$ or $D_6$ CS grids by a constant shift smaller in magnitude than $\eta_\mathrm{max}$. 
Thus the set of local environments that exist within the low-symmetry moir\'{e} cell will approach those in the symmetric cells as the twist angle decreases.
To examine the effect of lower symmetry, we performed calculations for a range of structures generated from an AA-stacked bilayer using different rotation centers, resulting in both high-symmetry and low-symmetry configurations (see Supplemental Material \cite{SM} for details).
We find that the cohesive energy per atom for low-symmetry structures is identical to that of their high-symmetry counterparts at small twist angles.
Furthermore, all configurations exhibit similar relaxation fields. This numerical agreement reflects the smooth and slowly varying nature of the moiré potential across the unit cell. By extension, we expect the same reasoning to apply to \textit{incommensurate} configurations, which can be thought of as consisting of many moiré cells with similar (but not identical) CS grids, resulting in an average interlayer hamiltonian $\mathcal{H}_\perp$ that closely matches that of a single commensurate cell.

\section{Conclusions}
The remarkable tunability of moiré materials arises from competing length scales and thus very large number of atoms in their unit cells, rather than from chemical composition. 
Their structural complexity makes symmetry analysis indispensable for studying these systems. 
In this work, we have shown that the atomic and electronic degrees of freedom in twisted bilayer graphene are traced to an interlayer hamiltonian with $D_6$ point group symmetry, regardless of the specific stacking geometry or choice of rotation center.  
This universality leads to identical atomic relaxation fields, electronic band structures, and wavefunctions. 
Although various configurations differ in total symmetry, they share the same local interlayer physics; this is particularly important when interpreting real experimental samples, which are often incommensurate. Our results form the foundation for constructing reliable interacting real-space lattice models that offer a promising alternative to momentum-space approaches; the latter often yield competing ground states that are difficult to disentangle~\cite{zhu2024weak}. An interesting future direction is to explore how universal symmetries manifest in multilayer moiré systems.

\section{Acknowledgements}
We thank Gayani N.~Pallewela and Hasan M.~Abdullah for helpful discussions. 
The authors acknowledge the US Army Research Office (ARO) MURI project under grant No.~W911NF-21-0147 and the Simons Foundation award No.~896626. MMEA acknowledges support from the Singapore 
Ministry of Education,
Research Centre of Excellence Institute for Functional Intelligence Materials. AZ is supported by the U.S. Department of Energy 
Computational Science Graduate Fellowship, Award Number DE-SC0025528.

\let\oldaddcontentsline\addcontentsline 
\renewcommand{\addcontentsline}[3]{} 

\bibliography{Bibliography}

\clearpage
\onecolumngrid
\let\section\originalsection  

\begin{center}
\textbf{\Large Supplemental Material}
\end{center}

\setcounter{equation}{0}
\setcounter{figure}{0}
\setcounter{table}{0}
\setcounter{page}{1}
\setcounter{secnumdepth}{2}
\makeatletter
\renewcommand{\thepage}{S\arabic{page}}
\renewcommand{\thesection}{S\arabic{section}}
\renewcommand{\theequation}{S\arabic{equation}}
\renewcommand{\thefigure}{S\arabic{figure}}
\renewcommand{\thetable}{S\arabic{table}}

\vspace{1cm}

\section{High-symmetry Twisted Configurations}

 Twisting two AA-stacked graphene layers about a point with vertically stacked atoms (Fig \ref{fig1}b, left) yields a structure with $D_3$ symmetry, while twisting about the center of vertically stacked hexagons (Fig \ref{fig1}a, left) results in $D_6$ symmetry. Conversely, twisting two AB-stacked layers about a position with vertically stacked atoms (Fig \ref{fig1}a, right) leads to a moir\'{e} cell with $D_6$ symmetry, though the $D_6$ axis is not located at the original twisting center.  Twisting about a point where a hexagon center in one layer is vertically aligned with an atom in the other layer (Fig \ref{fig1}b, right) results in $D_3$ symmetry. These combinations of stacking and rotation centers are summarized in Table~\ref{tabS1}. In the case of twisted transition metal dichalcogenides like  WSe$_2$, all four high-symmetry twist axes lead to $D_3$ symmetry, albeit with two distinct atomic arrangements. 

\begin{table}[h]
\centering
\renewcommand{\arraystretch}{1.5} 
    \begin{tabular}{c|c||c|c}
        \multicolumn{2}{c||}{AA stacking} & \multicolumn{2}{c}{AB stacking} \\
        \hline
        rotation center & symmetry & rotation center & symmetry \\
        \hline
        atom-atom   & $D_3$  & atom-atom & $D_6$ \\
       hex-hex & $D_6$  & hex-atom & $D_3$ \\
    \end{tabular}
\caption{
Point groups of twisted bilayer graphene configurations for different combinations of rotation center and initial untwisted vertical stacking. The rotation axis is perpendicular to both layers and passes either through a carbon site (``atom") or the geometric center of a ring of 6 atoms (``hex") in each layer.
 }
\label{tabS1}
\end{table}

\section{Low-symmetry Twisted Configurations}

In addition to the high-symmetry twisted configurations discussed in Table \ref{tabS1}, commensurate structures with lower symmetry can also be generated by first applying a twist and then shifting the layers—either rigidly shifting both layers together or shifting them independently. Equivalent results can also be obtained by first shifting the rotation center and then applying the twist.  In all cases, the twist is performed at commensurate angles matching those of the high-symmetry structures. This process generically yields commensurate structures with reduced symmetry.

\begin{figure*}[t]
\centering
\includegraphics[width=1.0\textwidth]{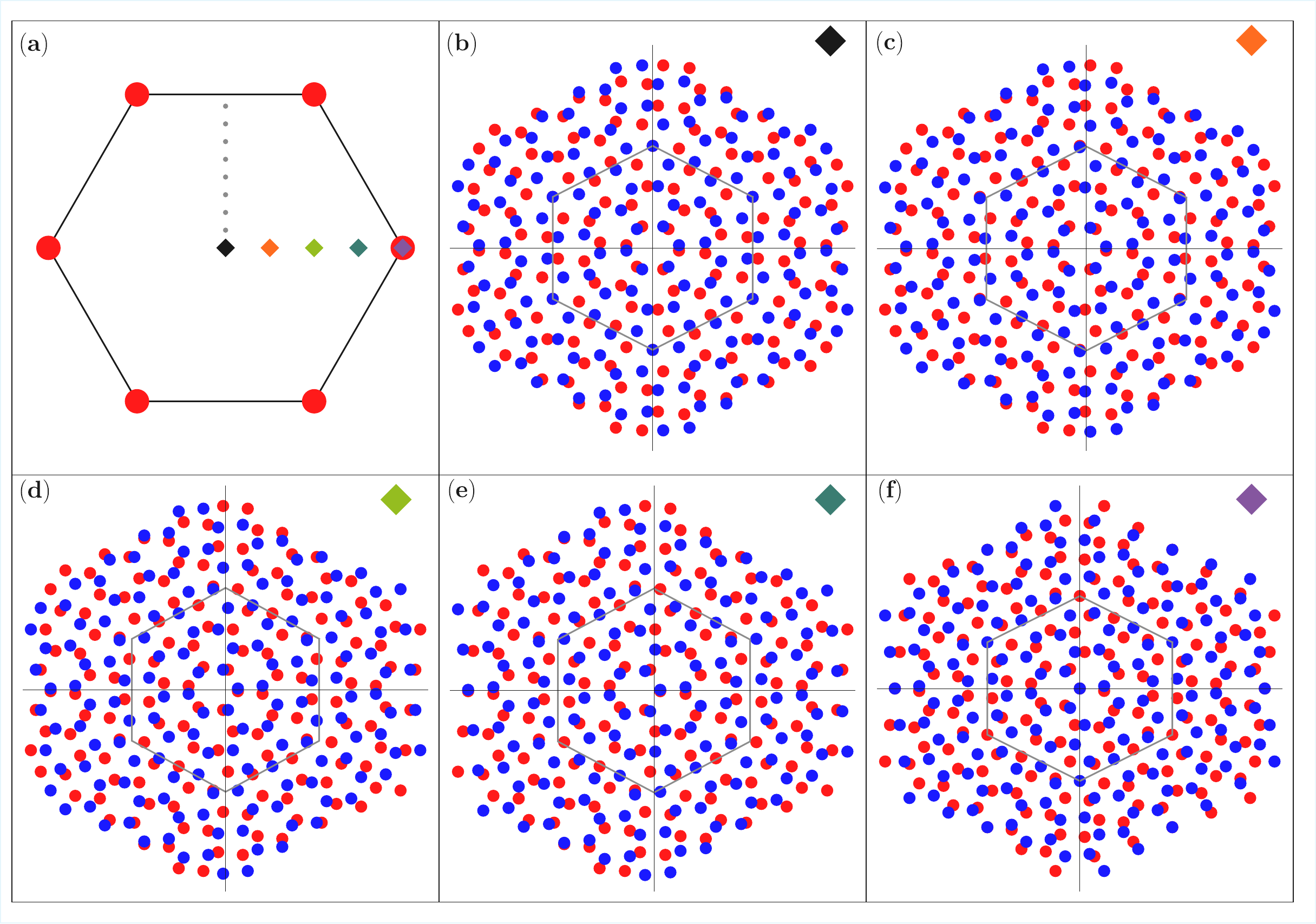}
\caption{Twisted bilayer configurations generated from different rotation centers.
(a) Schematic of the Wigner–Seitz cell for untwisted AA-stacked bilayer graphene with several candidate rotation centers marked by colored diamonds and gray dots. The red circles represent carbon atoms with the distance between them given by $a/\sqrt{3}$. Twisting about the black and purple diamonds yields high-symmetry configurations with $D_6$ and $D_3$ point groups, respectively. Twisting about intermediate positions of other marked centers results in lower-symmetry configurations.
(b–f) Real-space atomic structures corresponding to the rotation centers marked in panel (a). Red and blue dots denote atoms from the top and bottom layers, respectively. The gray hexagon outlines the Wigner–Seitz moiré cell. The structure evolves continuously as the rotation center is shifted with respect to the graphene lattice. While the overall pattern appears to rigidly shift, subtle internal rearrangements occur, reflecting the change in symmetry. A large twist angle of $13.2^\circ$ is used here for visual clarity.}
\label{figS1}
\end{figure*}

Figure \ref{figS1} illustrates a set of twisted configurations obtained by starting with an AA stacked bilayer and rigidly shifting both layers to the left so that the rotation center at the origin falls at different locations with respect to the carbon atoms. In Fig.~\ref{figS1}(b) the twist axis at the origin fall in the center of both carbon rings, yielding a high-symmetry $D_6$ structure. As the layers are shifted the rotation center moves through several intermediate low-symmetry locations, Figs.~\ref{figS1}(c–e), finally arriving at a pair of vertically stacked carbon atoms, Fig.~\ref{figS1}(f), which results in a structure with $D_3$ symmetry.

In addition to the configurations shown in Fig.~\ref{figS1}(b–f), one can shift the rotation center in other directions as well—an example is provided by the gray dots along the vertical axis in Fig.~\ref{figS1}(a). These alternative choices also yield commensurate structures with reduced symmetry. Despite their different atomic distributions, all such configurations exhibit identical cohesive energies per atom at small twist angles, reinforcing the idea that local stacking, rather than global symmetry, governs the interlayer energetics.

\section{Molecular Dynamics Simulations}

Molecular dynamics simulations were done using the Large-scale Atomic/Molecular Massively Parallel Simulator (\textsc{lammps}) code which employs classical interatomic force field models \cite{THOMPSON2022108171}. For twisted bilayer graphene, we use the Drip potential for interlayer interactions and the REBO potential for intralayer interactions \cite{brenner2002a,wen2018dihedral}. For twisted WSe$_2$ we use the KC potential for interlayer interactions and the SW potential for intralayer interactions with SW/mod style \cite{naik2019kolmogorov,jiang2015parametrization}.

\section{DFT Calculations}

Density functional theory (DFT) calculations were performed in the {\sc siesta} \cite{soler2002siesta} code, version 5.0, with norm-conserving \cite{normconserving} {\sc psml} pseudopotentials \cite{psml} obtained from Pseudo-Dojo \cite{pseudodojo}, and a basis of numerical atomic orbitals \cite{junquera2001numerical}.
In order to keep the calculations efficient, a basis of single-$\zeta$ polarized (SZP) orbitals optimized for graphene~\cite{bennett2025accurate} were used for all calculations, with $2s$, $2p$ orbitals for each carbon atom, and $3d$ polarization orbitals.

The PBE exchange-correlation functional \cite{pbe} was used in the calculations, and a DFT-D3 dispersion correction \cite{grimme2010consistent} was included to treat the long range interactions between the layers.
A single $k$-point was used in all calculations, and the real space grid was determined using an energy cutoff of 300 Ry.
The SCF was converged until the relative changes in the density matrix were below $10^{-4}$.

Twisted bilayer structures were generated for a twist angle of $\theta = 5.09^{\circ}$ (508 atoms).
Starting with two stacked, flat graphene layers separated by $d=3.35$ \AA, the atomic positions were relaxed until the forces on all atoms were below 1 meV/\AA.
The center of mass was fixed in the out-of-plane direction to prevent the bilayer from drifting from its initial position in the cell. After obtaining the relaxed geometries, the band structure was calculated along the path $\mathrm{K}-\Gamma-\mathrm{M}-\mathrm{K}$, with 10 points along each segment.

\subsection{Wavefunctions}

In {\sc siesta}, three-dimensional wavefunctions are given as linear combinations of the numerical atomic orbitals $\phi_i$:
\begin{equation}
    \Psi({\rv}) = \sum_i c_i\phi_i(\rv).
    \label{eq:wfsx-orbitals}
\end{equation}
To visualize the wavefunctions in two-dimensions, we first obtain modulus squared wavefunctions summed over the $z$ direction:
\begin{equation}
    \vert\Psi(x, y)\vert^2 = \sum_z
\vert\Psi(x, y, z)\vert^2.
\end{equation}
These squared wavefunctions have atomic-scale features. To extract and visualize the moir\'e-scale features, we perform a Gaussian convolution on the squared wavefunctions:
\begin{equation}
    \widetilde{\Psi}(x, y) \equiv (\vert\Psi\vert^2 * \mathcal{N})(x, y) = \sum_{x', y'} \vert \Psi(x', y')\vert^2 \mathcal{N}_{x,y}(x',y'),
\end{equation}
where $\mathcal{N}_{x,y}$ is a 2D symmetric Gaussian distribution centered at $(x, y)$ with covariance matrix $\sigma^2 I_{2 \times 2}$. We sum over $(x', y')$ within a $3\sigma$ radius of $(x, y)$ with grid spacing $\Delta x = \Delta y = 0.25$ \AA, and we set $\sigma = 2.5$ \AA for the moir\'e-scale wavefunctions $\widetilde{\Psi}(x, y)$ shown in Fig. \ref{fig4} in the main text. Lastly, the sublattice-projected wavefunctions are obtained by taking linear combinations of orbitals residing in the same sublattice ($L_1A$, $L_1B$, $L_2A$, or $L_2B$) in Eq. \ref{eq:wfsx-orbitals}.

Table \ref{tabS2} shows the symmetry representations of the wavefunctions analyzed using the atomic-scale DFT wavefunctions $\Psi({\rv})$.

\begin{table}[h]
\centering
\renewcommand{\arraystretch}{1.5} 
    \begin{tabular}{c||c|c}
        \multirow{2}{*}{Symmetry} & \multicolumn{2}{c}{Eigenvalue} \\ 
        \cline{2-3}
        & $D_3$ & $D_6$ \\
        \hline
        $C_{3}$  & $(1, 1), (1, 1)$ & $(1, 1), (1, 1)$ \\
        $\mathcal{M}_y$ & $(1, 1), (-1, -1)$ & $(1, 1), (-1, -1)$ \\
        $C_{2z}$  & $(1, -1), (1, -1)$ & $(1, -1), (1, -1)$ \\
    \end{tabular}
\caption{
Symmetry representations of the low-energy  bands of tBLG at the $\Gamma$ point. $\mathcal{M}_y$ is a $180^\circ$ rotation about an axis parallel to the layers that exchanges layers and sublattices. $C_{2z}$ is a $180^\circ$ rotation about an axis perpendicular to the layers that exchanges only sublattices within a layer. Numbers indicate eigenvalues of the sublattice-projected wavefunctions under each symmetry operation. Eigenvalues of wavefunctions in degenerate bands are grouped in parentheses.
 }
\label{tabS2}
\end{table}

\end{document}